\documentclass[lettersize,journal]{IEEEtran}
\usepackage{amsmath,amsfonts}
\usepackage{algorithmic}
\usepackage{algorithm}
\usepackage{array}
\usepackage[caption=false,font=normalsize,labelfont=sf,textfont=sf]{subfig}
\usepackage{textcomp}
\usepackage{stfloats}
\usepackage{url}
\usepackage{verbatim}
\usepackage{graphicx}
\usepackage{cite}
\usepackage{bm}
\usepackage{tabularx}
\usepackage{booktabs}
\usepackage{float}
\usepackage{color}
\usepackage{soul} 
\usepackage{flushend}
\usepackage{amssymb}

\newtheorem{assumption}{Assumption}

\newtheorem{proposition}{Proposition}

\newtheorem{definition}{Definition}

\newtheorem{remark}{Remark}
\newtheorem{fact}{Fact}
\usepackage{overpic}

\def\begquo{\begin{quote}}
\def\endquo{\end{quote}}
\def\begequarr{\begin{eqnarray}}
\def\endequarr{\end{eqnarray}}
\def\begequarrs{\begin{eqnarray*}}
\def\endequarrs{\end{eqnarray*}}
\def\begarr{\begin{array}}
\def\endarr{\end{array}}
\def\begequ{\begin{equation}}
\def\endequ{\end{equation}}
\def\lab{\label}
\def\begdes{\begin{description}}
\def\enddes{\end{description}}
\def\begenu{\begin{enumerate}}
\def\begite{\begin{itemize}}
\def\endite{\end{itemize}}
\def\endenu{\end{enumerate}}

\def\lef[{\left[\begin{array}}
\def\rig]{\end{array}\right]}
\def\qed{\hfill$\Box \Box \Box$}
\def\begcen{\begin{center}}
\def\endcen{\end{center}}
\def\begrem{\begin{remark}\rm}
\def\endrem{\end{remark}}
\def\begdef{\begin{definition}}
\def\enddef{\end{definition}}
\def\begpro{\begin{proposition}}
\def\endpro{\end{proposition}}
\def\begfac{\begin{fact}}
\def\endfac{\end{fact}}
\def\begass{\begin{assumption}}
\def\endass{\end{assumption}}

\def\begmat#1{\begin{bmatrix}#1\end{bmatrix}}
\def\begali#1{\begin{align}{#1}\end{align}}
\def\begalis#1{\begin{align*}{#1}\end{align*}}

\def\calp{{\bf p}}

\def\liminf{\lim_{t \to \infty}}

\def\L2e{{\cal L}_{2e}}

\def\rea{\mathbb{R}}
\def\intnum{\mathbb{N}}

\def\bfthe{{\boldsymbol\theta}}

\def\bfe{{\bf e}}

\def\bfome{{\boldsymbol{\Omega}}}

\def\bfy{{\bf Y}}
\def\ya{y_{1,2}}
\def\yb{y_{3,4}}
\def\yc{y_{5,6}}

\DeclareMathOperator{\atantwo}{atan2}

\def\y{{y}}

\begin{document}

\title{A New Adaptive  Phase-locked Loop for Synchronization of a Grid-Connected Voltage Source Converter: Simulation and Experimental Results}

\author{
	\vskip 1em
	{
		Wei He,
	     Jiacheng Yan,
          Romeo Ortega,
          Daniele Zonetti
	}
	
	\thanks{
		
		{

			Wei He and Jiacheng Yan are with Jiangsu Collaborative Innovation Center of Atmospheric
Environment and Equipment Technology (CICAEET), Nanjing University
			of Information Science and Technology, Nanjing 210044, China.

               Romeo Ortega is with the Departamento Acad$\grave{e}$mico de Sistemas Digitales, ITAM, Rio Hondo 1, Col. Progreso Tizapan, 01080 Ciudad de M$\grave{e}$xico, Mexico.
			
			Daniele Zonetti is with the Centre d'Innovaci$\acute{o}$ Tecnol$\grave{o}$gica en Convertidors Est$\acute{a}$tics i Accionaments, Departament
d'Enginyeria El$\acute{e}$ctrica, Universitat Polit$\acute{e}$cnica de Catalunya, Barcelona 08028, Spain.
		}
	}
}

\markboth{}%
{Shell \MakeLowercase{\textit{et al.}}: A Sample Article Using IEEEtran.cls for IEEE Journals}


\maketitle

\begin{abstract}
	In \cite{ZONetal_aca} a new adaptive phase-locked loop scheme for synchronization of a grid connected voltage source converter with guaranteed (almost) global stability properties was reported. To guarantee
a suitable synchronization with the angle of the three-phase grid voltage we design an adaptive observer for
such a signal requiring measurements only at the point of common coupling.  In this paper we present some simulation and experimental illustration of the excellent performance of the proposed solution.
\end{abstract}


\section{Introduction}
\label{sec1}
The landscape of energy generation is undergoing unprecedented transformations, as renewable energy sources (RES) become increasingly prominent in modern power grids. A key role in this transformation is played by voltage source converters (VSCs), which enable the integration of RES, such as photovoltaic panels and wind turbines, into the existing grid \cite{BOSEbook,HOFetal,mirafzal2014survey,cecati2022state}. VSCs exhibit distinct characteristics from conventional synchronous generators---which are the devices that have historically dominated power grids---and possess a notable advantage in terms of enhanced controllability. However, capitalizing on this advantage demands the formulation of robust control strategies that can effectively withstand significant perturbations generated by variations in power supply and demand, transient disturbances, and intermittent RES.

These perturbations typically manifest as substantial voltage and frequency fluctuations at the Point of Common Coupling (PCC), which become more relevant in power grids characterized by a low level of inertia and low short-circuit capabilities. Although there is an interest in developing a unified approach to deal with both scenarios, there exist many practical situations where the grid is characterized by a relatively high level of inertia, yet low short-circuit capabilities. These include low-voltage distribution systems characterized by a large share of constant power loads or where the cables used for transmission have a small section \cite{WUetal}; high-voltage transmission systems, whereas large transformers are needed to step up the voltage to facilitate transmission \cite{EGEetal}. It is common to refer to grids with short-circuit capabilities as \textit{weak} grids, irrespectively from their level of inertia.

A common approach for control design in grid-connected VSC is the use of the so-called $\tt{dq}$ controllers, i.e., control architectures that structurally rely on a suitably defined {synchronous} rotating frame, whose reference angle dictates the mode of operation of the VSC \cite{DAVetal}. Whenever the reference angle is selected to synchronize up to a given phase shift (\textit{follow}), with the grid angle, the VSC is said to operate in \textit{grid-following} mode; whenever it is designed (\textit{formed}) via a suitably defined frequency control loop, the VSC is said to operate in \textit{grid-forming} mode \cite{MILetal2018}. While for the grid-forming mode of operation a variety of strategy have been proposed requiring only knowledge of the grid frequency, for the grid-following mode information about the grid angle remains critical. Since this information remains inaccessible to the system operator, an appropriate algorithm must be developed to estimate this signal. Traditional design relies on the use of phase-locked loops (PLLs) mechanisms\cite{ABR}. Nevertheless, in weak grids subject to substantial voltage fluctuations, ensuring a stable reference angle for the PLL becomes considerably complex, since the varying voltage levels during disturbances can cause fluctuations in the frequency and phase angle estimates, disrupting the ability of the PLL to provide a synchronous reference frame. Consequently, hierarchical controllers that relies on this rotating frame may be prone to instability.

In literature numerous solutions to address the challenges posed by weak grids have been proposed. A possible approach involves refining the operation of the PLL to maintain stability. This requires careful adjustments of the PLL and $\tt{dq}$ controller gains to ensure stability. Traditionally, the power electronics community addressed this problem by analysing the small-signal model of the grid-connected VSC, see \cite{ZHOU} and references therein for an overview. This strategy has inherent limitations as it is either built upon a small-signal analysis of the system. Consequently, it remains valid only in close proximity to equilibrium conditions. More recently, Lyapunov-based analysis based on a time-scale separation between the current controller dynamics and the PLL have been proposed \cite{MANetal}, \cite{ZHAOetal}. While these analysis provide further insights on the selection of the PLL gains, the results lead to conservative, {local} stability certificates whose applicability is highly questionable. Alternative approaches are based on the re-design of outer control loops, which in normal conditions would operate at a much slower time-scale \cite{WANGetal}. In these approaches, the outer loops are brought closer in timescale to the PLL's operation. The purpose is to step in and enforce stability when the PLL is susceptible to failure due to voltage fluctuations. These strategies often rely on intuitive power system knowledge and insights. Yet, the overall effectiveness and reliability of these approaches raise questions about their overall practical applicability.

In a recent paper \cite{ZONetal_aca}, we presented an adaptive solution for the synchronization of a grid-connected VSC, enhancing the standard PLL design based on a $\tt dq$ transformation of the system dynamics. More precisely, we proposed to complement the conventional synchronization scheme with an \textit{observer} that reconstructs the $\tt dq $ voltage behind the grid impedance---an information that is next provided to a PLL to ensure synchronization. To solve this problem at hand we proposed in that paper a novel approach that combines recently developed and more conventional techniques in estimation theory: the generalized parameter-based estimation observer (GPEBO) technique is used to derive a linear regressor equation (LRE) needed for the estimation of the grid voltage; a classical least-squares estimator\cite{LJUbook,SASBODbook,TAObook} (LS)---improved with a forgetting factor (FF) mechanism---is used to estimate such a voltage; a conventional PLL is applied to the estimated voltage to recover the grid phase.   Significantly, we showed that under some classical \textit{persistency of excitation} assumption, the proposed solution guarantees almost global convergence of the system's solutions---a result that, as already mentioned, should be contrasted with local stability results available in literature.

In this context, the present paper aims thus to substantiate the theoretical results obtained in \cite{ZONetal_aca} with practical experiments. The remainder of the paper is organized as follows. In Section \ref{sec2}, to place our contribution in context, we first review the standard solutions to the grid synchronization problem. The main result is given in Section  \ref{sec3}, where we present our proposed observer-based approach. Some illustrative simulation results are presented in Section \ref{sec4} while experimental results are given in Section \ref{sec5}. We wrap-up the paper with concluding remarks and future research in Section \ref{sec6}.

\noindent {\bf Notation.} The symbols $\intnum_{+}$, $\rea_+$ denote  the sets of positive integers and positive real numbers respectively, and $\mathbb S$ is the unit circle in $\rea^2$.  Given $n \in \intnum_{+}, q \in \intnum_{+}$, $\mathbb{I}_n$ is the $n \times n$ identity matrix and ${\bf 0}_{n\times q}$  is an $n \times q$ matrix of zeros. $\bfe_q \in \rea^n$ denotes the $q$-th vector of the $n$-dimensional Euclidean basis. For $x \in \rea^n$, we denote the square of the Euclidean norm as $|x|^2:=x^\top x$.  Given ${J}:=\begmat{0&-1\\1&0}$ and $\alpha\in \rea$ we define the rotation matrix as
$e^{{J}\alpha}=\begin{bmatrix}
       \cos(\alpha)&-\sin(\alpha)\\
       \sin(\alpha)&\cos(\alpha)
       \end{bmatrix}.$
$\atantwo:\mathbb S\rightarrow [-\pi\;\pi)$ denotes the 2-arguments arctangent function. Given a differentiable signal $u(t)$, we define the derivative operator $\calp^i[u]=:{d^iu(t)\over dt^i}$ and denote the action of a linear time-invariant (LTI) filter $F(\calp) \in \rea(\calp)$  as $F[u]$.
%
 \section{Grid Synchronization Problem and Standard PLL Solutions}
\label{sec2}
\begin{figure*}
  \centering
  \includegraphics[width=0.8\textwidth]{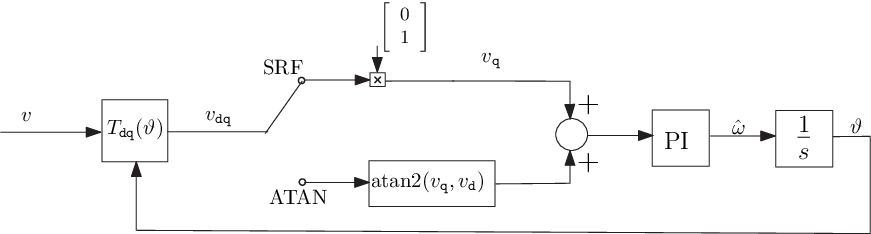}
  \caption{Block diagram representation of the SRF- and ATAN-PLL schemes. }
  \label{figa}
\end{figure*}

%
As indicated in the Introduction, the problem we address in this paper is the estimation of the angle of the grid voltage $v_g(t) \in\mathbb{R}^3$, which is in general \textit{not measurable}, but assumed to be of the form
\begin{equation}
\label{vg}
v_{g} = V_g \begin{bmatrix}
\sin(\omega t)\\
\sin(\omega t-\frac{2}{3}\pi)\\
\sin(\omega t+\frac{2}{3}\pi)
\end{bmatrix},
\end{equation}
where  $V_g\in\mathbb{R}_+$ and $\omega\in\mathbb{R}_+$  denote the respectively the voltage amplitude and frequency of the grid, which are \textit{both unknown}. {A common assumption is that the grid voltage $v_g$  only slightly differs from the voltage $v$ at the PCC, which is \textit{measurable}. Hence, $v\approx v_g$, meaning that the grid voltage can be considered as a signal \textit{known} with good accuracy. Based on this assumption,} the standard approach to accomplish the estimation task is to implement a PI controller that processes a {\em phase detector}, which is a signal containing information of the grid angle---this system is usually called a PLL, see \cite{CHU,COLbook,TEObook,YAZIRAbook} for tutorial introductions. Two widely popular schemes for the generation of the phase detector are the so-called synchronous reference frame (SRF) and the arctangent (ATAN)-PLL. Both of them rely on the  transformation of the voltage \eqref{vg} to ${\tt dq}$ coordinates via the well-known  matrix rotation $T_{\tt dq}$  \cite[eq. (2.3)]{SCHetal} using a (designer chosen) rotation angle $\vartheta$---which plays the role of an {\em estimate} of the grid phase.  That is, we define the new signals
$$
v_{\tt dq}:=T_{\tt dq}\left(\vartheta - {\pi \over 2}\right)v
$$
that, in view of the aforementioned assumption, yields the rotated grid voltage
\begequ
v_{\tt dq}\approx v_{g\tt dq}=V_g\begmat{\cos(\phi) \\ \sin(\phi)},
\label{vgdq0}
\endequ
where we defined the {\textit{error signal}} $\phi(t)\in\mathbb R$ as
 \begin{equation}
 \lab{del}
  \phi:=\vartheta-\omega t.
 \end{equation}
In the SRF-PLL the phase detector is the second component of the vector $v_{\tt dq}$,  and this signal is fed to the PI. The output of the PI, is viewed as an estimate of the unknown grid frequency, denoted $\hat \omega$. In the ATAN-PLL the phase detector is generated computing the arctangent of the error signal $\phi$, via the $\atantwo$ function, as shown in Fig. \ref{figa}.  In both schemes, the output of the PI is  integrated to generate an estimate of the grid phase $\vartheta$, that is used as the argument of the rotation matrix---see Fig. \ref{figa}.    Interestingly, in spite of  the highly nonlinear dynamical systems that describe their behavior, it is possible to show {\em almost global stability} of both PLLs \cite{MANetal,RAN,ZONetal_ecc}. {Unfortunately, the assumption of negligible difference between the grid voltage and the voltage at the PCC falls short in the case of weak grids, posing questions about the ability of the PLL to guarantee stability in these conditions.}

%
 \section{An Observer-based Synchronization Method}
\label{sec3}
%
Similarly to the SRF- or ATAN-PLL schemes described above, the adaptive PLL that we consider in this paper, which was proposed in \cite{ZONetal_aca}, also implements a rotation of the grid voltage as \eqref{vgdq} and uses the PI+integration feedback structure of Fig. \ref{figa}, but {\em generates} the phase detector in a radically different way. Indeed, the problem of generation of an estimate of the error signal $\phi$ is posed as an {\em adaptive observer} design task. Towards this end, we look at the dynamics of the whole system, comprising  a three-phase, balanced two-level VSC interfaced to the balanced AC grid---modeled using the Thevenin equivalent (TE) circuit. The  overall model of the system is given as
  \begali{
  \nonumber
  L_g\dot i_g &=-r_gi_g+v -v_g\\
\nonumber
L\dot i &=-r i+V_\mathrm{dc}m-v\\
  C\dot{v} &= -i_g+i,
\label{sys_abc}
}
where $v_g(t) \in\mathbb{R}^3$ is the voltage of the grid assumed to be of the form \eqref{vg}, with $V_g\in\mathbb{R}_+$ and $\omega\in\mathbb{R}_+$   \textit{unknown}. The \textit{measurable} state variables $i_g(t)\in\mathbb{R}^3$, $i(t)\in\mathbb{R}^3$  and $v(t)\in\mathbb{R}^3$ denote the current of the grid, the current through the phase reactor and the AC voltage at the PCC, respectively, $V_\mathrm{dc}\in\mathbb{R}_+$ is the {\em known constant} DC voltage and $m(t)\in \mathbb{R}^3$ are the modulation indices. The positive parameters $L$, $r$, $L_g$, $r_g$ and $C$ are, respectively, the inductance and resistance of the phase reactor, the inductance and resistance associated to the grid and the capacitance of the filter---with $L_g$ and $r_g$ \textit{known}.

In \cite{ZONetal_aca} it is proposed to represent the system dynamics \eqref{sys_abc} in a  $\tt dq $ reference frame with a transformation angle $\vartheta$, generated by
$$
\dot \vartheta=u_1
$$
with $u_1(t) \in \rea$ a signal {\em to be defined} such that
\begequ
\lab{limvarthe}
\liminf \vartheta(t)=\omega t + \phi^\mathrm{ref}
\endequ
is ensured---where, for the sake a generality, we have added the constant $\phi^\mathrm{ref}$, which is a user selected constant phase shift. That is, we define the new signals
$$
(\cdot)_{\tt dq}:=T_{\tt dq}\left(\vartheta - {\pi \over 2}\right)(\cdot)_{\tt abc},
$$
with the definition of $T_{\tt dq}$ given in \cite[eq. (2.3)]{SCHetal}. This operation yields the system dynamics
\begin{equation}
\begin{aligned}
    \begin{bmatrix}
    L_g\dot{i}_{g\tt{dq}}\\
    C\dot{v}_{\tt{dq}}\\
    L\dot{i}_{\tt{dq}}
    \end{bmatrix}=&
    \begin{bmatrix}
     -r_g\mathbb{I}_2&\mathbb I_2&{\bf 0}_{2\times 2}\\
   -\mathbb I_2& {\bf 0}_{2\times 2} &\mathbb{I}_2\\
    {\bf 0}_{2\times 2}&-\mathbb{I}_2& -r\mathbb{I}_2
    \end{bmatrix}
    \begin{bmatrix}
    i_{g\tt{dq}}\\ v_{\tt{dq}}\\ {i}_{\tt{dq}}
    \end{bmatrix}\\&+\begin{bmatrix}
    L_gJi_{g\tt{dq}} & {\bf 0}_{2\times 2}\\
    CJv_{\tt{dq}} & {\bf 0}_{2\times 2} \\
    {LJi_{\tt{dq}}} & \mathbb{I}_2
    \end{bmatrix} \begmat{u_1 \\ u_2 \\ u_3}
    +\begin{bmatrix}
    - v_{g\tt dq} \\{\bf 0}_{2 \times 1}\\{\bf 0}_{2 \times 1}
    \end{bmatrix}\label{sys-dq}
\end{aligned}
\end{equation}
 where we  defined the new control signal
\begin{equation*}\label{eq:vdc-vgdq}
\begmat{u_2 \\ u_3}  :=V_\mathrm{dc}m_{\tt{dq}},
\end{equation*}
and the rotated grid voltage \eqref{vgdq0} that, to simplify the future calculations, we rewrite as
\begequ
v_{g\tt dq}:=V_g e^{{J}\phi}\mathbf{e}_1,
\label{vgdq}
\endequ
where $\phi\in\mathbb R$ is the \textit{error signal} defined in \eqref{del}. This signal, which is of course {\em unknown}, satisfies the dynamic equation
\begin{equation*}\label{eq:dot-delta}
    \dot\phi=-\omega+u_1.
\end{equation*}
Moreover, the signal $v_{g\tt dq}$ given in \eqref{vgdq}, that we treat as an \textit{unmeasurable state}, satisfies the differential equation
\begin{equation*}\label{eq:dot-vg}
\begin{aligned}
    \dot v_{g\tt dq}&=\dot\phi {J} v_{g\tt dq}\\
                    &=-\omega {J} v_{g\tt dq}+u_1  {J} v_{g\tt dq}.
    \end{aligned}
\end{equation*}
Therefore, considering \eqref{del}, we see that the synchronization objective \eqref{limvarthe} can be recast as follows.
\begenu
 \item[{\bf SO}] Select the ``control signal" $ u_1$ to ensure that the unmeasurable state $v_{g\tt dq}$ satisfies
\begequ
\label{synobj}
\liminf v_{g\tt dq}(t)=V_ge^{{J}{\phi}^\mathrm{ref}}\mathbf{e}_1.
\endequ
\endenu
To solve this problem we design an {\em adaptive observer} for the state $v_{g\tt dq}$, which is given in  Proposition \ref{pro1} below.

To streamline the presentation of this result we find convenient to rewrite the system   using the standard control theory notation,  defining the {\em unmeasurable} state $x(t) \in \rea^2$, the {\em measurable} state $y(t) \in \rea^6$ and the input vector $u(t) \in \rea^3$ as
$$
{x:={1 \over L_g}v_{g\tt dq},}\quad
y:= \begmat{
    i_{g\tt{dq}}\\ v_{\tt{dq}}\\ {i}_{\tt{dq}}},
$$
respectively, and write the system dynamics as
\begin{equation}
\begin{aligned}
\dot y =&
    \begmat{
     -{r_g \over L_g}\mathbb{I}_2&{1 \over L_g}\mathbb I_2&{\bf 0}_{2\times 2}\\
   -{1 \over C}\mathbb I_2& {\bf 0}_{2\times 2} &{1 \over C}\mathbb{I}_2\\
    {\bf 0}_{2\times 2}&-{1 \over L}\mathbb{I}_2& -{r \over L}\mathbb{I}_2
    }y+    \begmat{
    { J} \ya & {\bf 0}_{2\times 2}\\
    { J} \yb & {\bf 0}_{2\times 2} \\
    { J} \yc & {1 \over L}\mathbb{I}_2
    }u
    \\&+\begmat{
    -x\\{\bf 0}_{2 \times 1}\\{\bf 0}_{2 \times 1}
    },
 \label{sys}
\end{aligned}
\end{equation}
where the unknown state $x$, to be reconstructed, satisfies the differential equation
\begequ
\label{sysx}
{\dot x}  {=-\omega { J} x+u_1 { J} x}.
\endequ

The proposition below describes in detail the structure of the proposed adaptive PLL and enunciates its stability properties. Its construction is based on GPEBO technique.

\begin{proposition}
\lab{pro1}\em
Consider the grid-connected VSC system  \eqref{sys_abc} and its $\tt dq$ representation \eqref{sys}, \eqref{sysx}. Define the dynamic extension
\begalis{
\dot z &= u_1{J}z-\frac{r_g}{L_g}{{J}}\begmat{y_{1}\\y_2} + \frac{1}{L_g}{{J}} \begmat{y_{3}\\y_4}\\
\dot \Phi & = u_1{J} \Phi,\;\Phi(0)=\mathbb{I}_2,
}
and introduce the signals
\begin{align*}
\bfy &:= \calp F(\calp)\Big[\begmat{y_{1}\\y_2}\Big] - F(\calp)\Big[\Big(u_1 {J} -\frac{r_g}{L_g} \mathbb I_2\Big)\begmat{y_{1}\\y_2} + \frac{1}{L_g}\begmat{y_{3}\\y_4} \Big]\\
\bfome   &:= -F(\calp)\Big[ \begmat{-z+{J} \begmat{y_{1}\\y_2} & \vdots & \Phi } \Big],
\end{align*}
where $F(\calp)={\lambda \over \calp+\lambda},\;\lambda>0$, is an LTI filter. Define the estimate of the state $x$ as
\begequ
\label{hatxi}
\hat x =\begmat{-z+{J} \begmat{y_{1}\\y_2} & \vdots & \Phi}\hat \bfthe,
\endequ
where $\hat \bfthe$ is generated by the least-squares parameter estimator
\begin{equation}\label{intestt1}
\dot{\hat \bfthe }    =\alpha F   \bfome^\top    (\bfy  -\bfome    \hat\bfthe  ),\; \hat\bfthe(0) \in \rea^3
\end{equation}		
\begin{equation}\label{phit1}
\dot {F}   =\begin{cases}
 -\alpha F   \bfome^\top    \bfome     F   {+ \beta F   } & \mbox{if} \; \|F  \| \leq M   \\
 0 & \mbox{otherwise}
 \end{cases}
\end{equation}
with  $F(0)={1 \over f_0} I_3$ and tuning gains $\alpha>0,\;f_0>0,\; {\beta \geq 0}$ and $M>0$. Define the {\em adaptive PLL} as
\begali{
\nonumber
\dot x_c &= e_\phi(\hat x) \\
\label{adapi}
u_{1} &=-K_{P} e_\phi(\hat x) -K_{I} x_c,
}
with gains $K_{P}>0$, $K_{I}>0$ and the {\em phase detector} $e_\phi(\hat x)$ chosen in either one of the forms
\begin{equation}\label{eq:PD}
\begin{aligned}
e_{\phi}(\hat x)&=-\cos({\phi}^\mathrm{ref})\hat x_2+\sin({\phi}^\mathrm{ref})\hat x_1,\\
\mathrm{or}\quad
e_{\phi}(\hat x)&=-\atantwo(\hat x_2,\hat x_1)+{\phi}^\mathrm{ref},
\end{aligned}
\end{equation}
corresponding (with $\hat x=x$) to an SRF-PLL or ATAN-PLL, respectively.\footnote{Conventional SRF- and ATAN-PLL aligned with the $\tt q$ axis can be recovered by picking ${\phi}^\mathrm{ref}=0$.}  Assume the vector $\bfome$ is {\em persistently exciting}, that is there exists constants $C_c>0$ and $T>0$ such that
	\begalis{
		&\int_{t}^{t+T} \bfome  ^\top(s) \bfome  (s)  ds \ge C_c I_3,\;\forall t \geq 0.
	}
Under these conditions, the synchronization objective \eqref{synobj} is achieved for \textit{almost} all system and controller initial conditions---guaranteeing that all signals remain bounded.	
\qed
\end{proposition}

Note that the signal $m_{\tt dq}$ is designed as a classical PI {\tt dq} current controller given in \cite[(27)]{ZONetal_aca}, which is omitted here. The detailed proof of above proposition is given in \cite[Section III]{ZONetal_aca}. We omit them here.

\begin{figure}[H]
  \centering
  \includegraphics[scale=0.8]{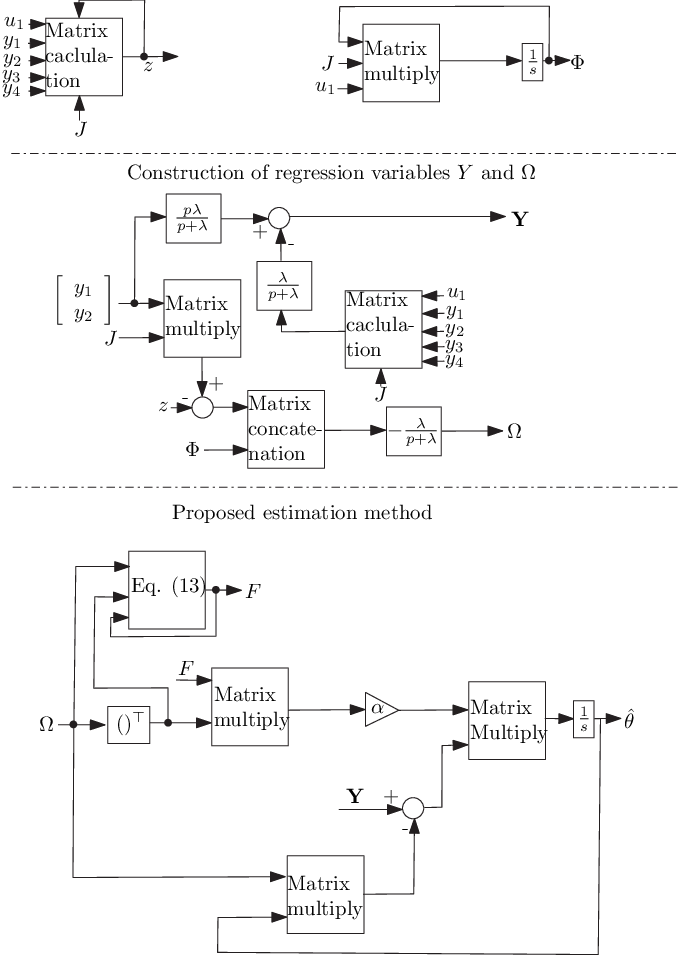}
  \vspace{0cm}
  \caption{The structure of the proposed estimation method.} \label{observer}
\end{figure}

\begin{figure*}
  \centering
  \includegraphics[scale=0.8]{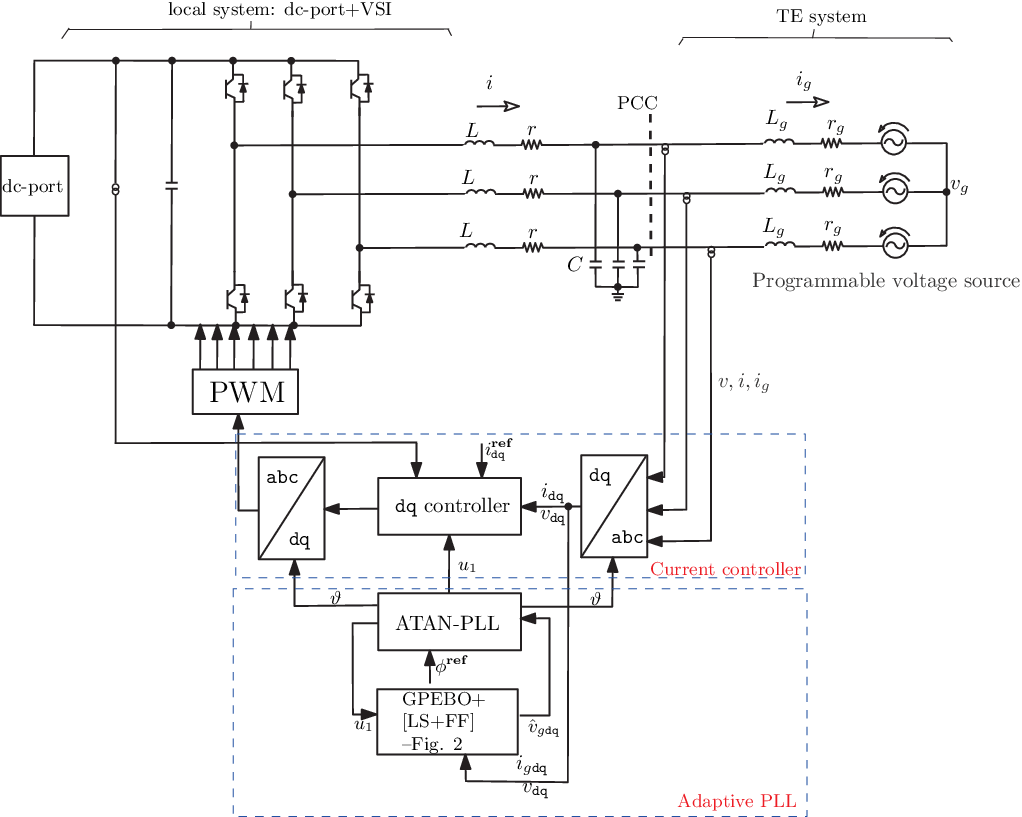}
  \vspace{0cm}
  \caption{Block diagram representation of the system \eqref{sys} and the proposed control scheme.}\label{fig0}
\end{figure*}

The structure of the proposed observer is shown in Fig.\ref{observer}. Besides,
a block diagram representation of the whole system, including the proposed adaptive PLL is given in Fig. \ref{fig0}.

%
\section{Simulation and experimental results}


\subsection{Simulation results}\label{sec4}
To validate the theoretical results we consider $P^\mathrm{ref}$=600 W, voltage reference $V^ \mathrm{ref}$=380 V, and the frequency $f=50$ Hz. Then, $\omega=2 \pi f$. The experimental parameters are shown in Tabs. \ref{tab1} and \ref{tab2}.

Based on the classical power flow equations, references for the $\tt{dq}$ currents and phase shift reference $\phi^\mathrm{ref}$---which are provided to the $\tt{dq}$ controller and PLL respectively---are calculated from the active power reference $P^\mathrm{ref}$  and voltage reference $V^ \mathrm{ref}$. It is noted that the calculation of power flow equations is done under nominal conditions in the simulation and experiment studies. This is to say that the parameters shown in Tabs. \ref{tab1} and \ref{tab2} are adopted for the calculation of power flow equations and the calculation results for the references will be fixed for different test scenarios. In the design we consider the current controller, together with the adaptive PLL where the estimated parameters $\mathbf{\hat \theta}$ are generated via the observer.  The gains of the adaptive PLL are set to $K_{P\phi} = 200, K_{I\phi} = 5000$, while the gains of the current controller are set, for both the $\tt{d}$ and ${\tt{q}}$ axis, to $K_P = 1250, K_I = 50000$. The gains of the LS+FF are, on the other hand, set to $\lambda=1000, \alpha =600, \beta = 500, M = 100, f_0 = 1.$
\begin{table}[!ht]
\caption{VSC Parameters.}
\centering
\begin{tabular}{lccc}
\hline
\hline
~Parameter~~& {Value} \\
\hline
~$f$~~~ & $50~\text{Hz}$ \\
~$C$~~ & $4.6$~$\mu F$ \\
~$L$~~ & $9.5$~mH \\
~$r$~~~ & $0.64$~$\Omega$\\
\hline
\hline
\end{tabular}
\label{tab1}
\end{table}
\begin{table}[!ht]
\caption{Variable power Source.}
\centering
\begin{tabular}{lccc}
\hline\hline
~Parameter~~& {Value} \\
\hline
~$L_g$~~ & $282$ mH \\
~$r_g$~~~ & $12.8$~$\Omega$ \\
~$V_g$~~~ & $310.2687~\text{V}$ \\
\hline\hline
\end{tabular}
\label{tab2}
\end{table}

\begin{figure}
  \centering
  \includegraphics[scale=0.6]{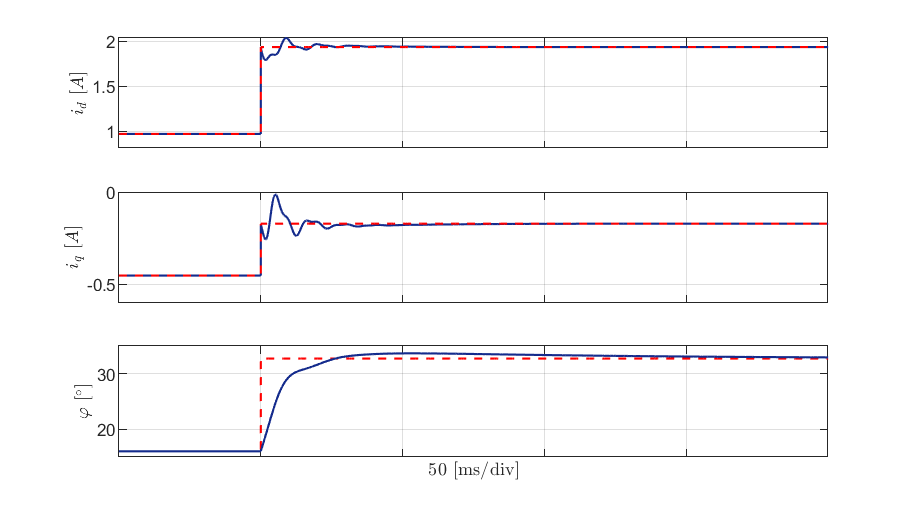}
  \caption{Responses of $\tt{dq}$ currents and angle $\varphi$, facing an active power step variation from $P_1^\mathrm{ref}$ to $P_2^\mathrm{ref}$.}\label{p-step}
\end{figure}

\begin{figure}
  \centering
  \includegraphics[scale=0.6]{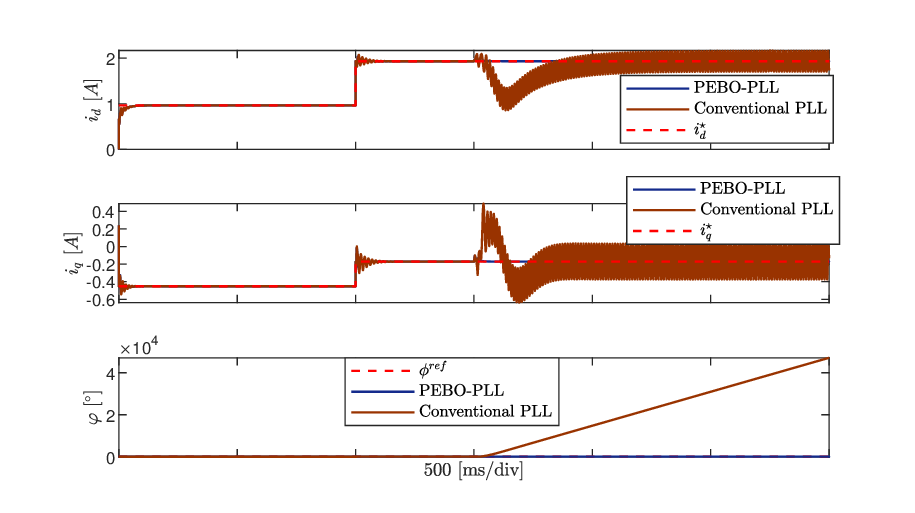}
  \caption{Responses of $\tt{dq}$ currents and angle $\varphi$, facing an active power step variation from $P_1^\mathrm{ref}$ to $P_2^\mathrm{ref}$ and a step change in $V_g$ under both proposed PEBO-PLL and conventional PLL schemes.}\label{comparison}
\end{figure}

{
We evaluate both nominal and perturbed scenarios, illustrating the responses of the directly controlled variables, i.e. the $\tt{dq}$ currents $i_{\tt{dq}}$ and of the phase shift $\phi = \vartheta - \omega t$. In the simulation and experiment results, we will show the angle $\varphi$ (degree) corresponding to the $\phi$ (radian), where $\varphi=\frac{180\phi}{\pi}$.

Firstly, we evaluate the ability of the proposed solution PEBO-ATAN-PLL to track variations in the active power profile, while keeping the voltage at the nominal value $V^\mathrm{ref}=380$ V. In particular, we consider a variation in the active power reference from $P_1^\mathrm{ref}=300$ W to $P_2^\mathrm{ref}=600$ W (from half to full rated value).
The $\tt dq$ currents $i_{\tt{dq}}$ and the angle $\varphi$ responses are illustrated in Fig. \ref{p-step}. It is observed that the corresponding references are tracked within $200$ ms.} {Besides, we carry out a comparison between the proposed adaptive PEBO-ATAN-PLL and conventionally non-adaptive ATAN-PLL. The gains are same for two schemes. The simulation result is shown in Fig. \ref{comparison}. It is seen that the angle $\varphi$ is unstable and ${\tt dq}$ currents responses contain obvious oscillation in the voltage $V_g$ with a step change in at $t=1.5$ s for conventional PLL, which shows a poor control performance.}

\begin{figure}
  \centering
  \includegraphics[scale=0.6]{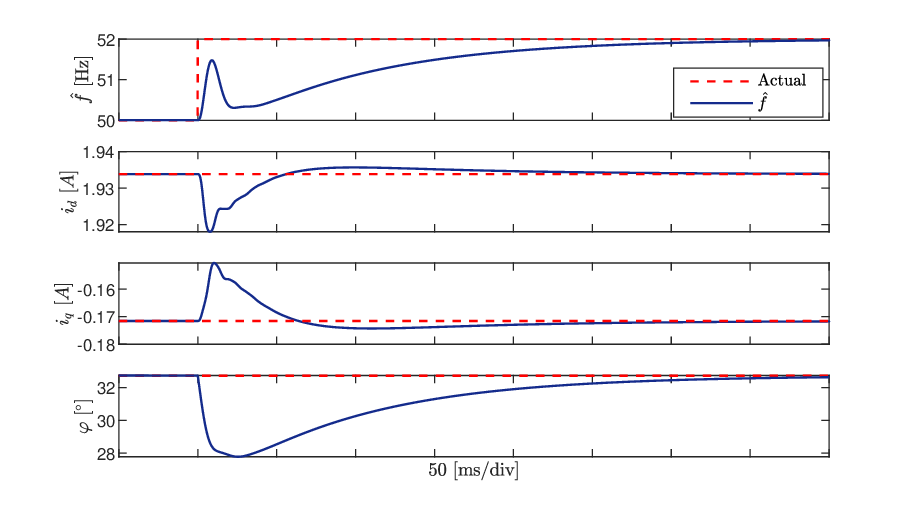}
  \vspace{-1cm}
  \caption{Responses of {\tt dq} currents and angle $\varphi$, facing a frequency step variation of $2$ Hz.}\label{w-step-id}
\end{figure}

\begin{figure}
  \centering
  \includegraphics[scale=0.6]{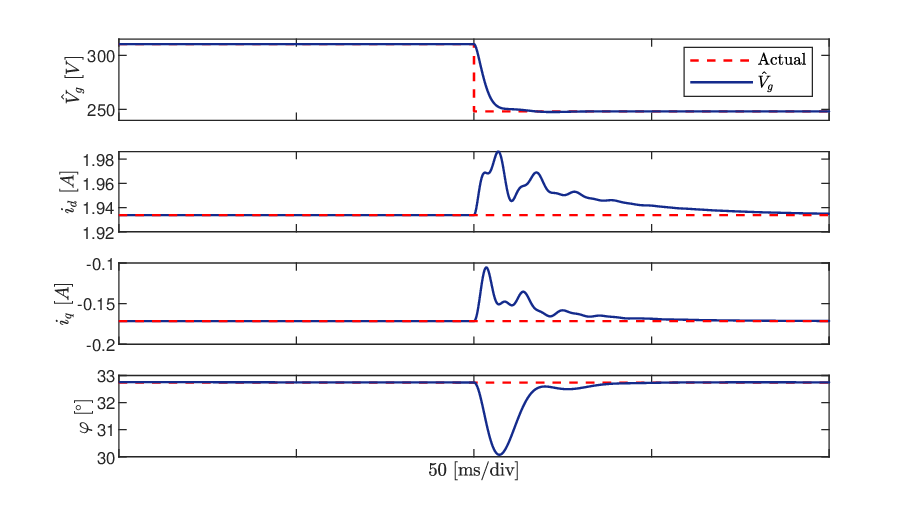}
  \vspace{-1cm}
  \caption{Responses of {\tt dq} currents and angle $\varphi$, facing a voltage amplitude step variation of $20 \%$.}\label{Vgg-step}
\end{figure}

{For the validation in perturbed conditions, the active power is supposed to be equal to the VSC rated value of $600$ W, maintaining the voltage to the nominal value $V^\mathrm{ref}=380$ V. Then, an abrupt variation of the frequency $f$, stepping up from $50$ Hz to $52$ Hz, is supposed to occur but for the calculation of power flow equations it has always been $50$ Hz. Hence, the references are unchanged when a step change in $f$ is considered. It can be observed from Fig. \ref{w-step-id} that the frequency variation is properly tracked, preserving stability of the overall system, with negligible fluctuations in the responses of the ${\tt dq}$ currents $i_{\tt dq}$. Similarly, we illustrate performances of the proposed solution in presence of an abrupt $20\%$ variation of the grid voltage amplitude $V_g$, stepping up from $310.2687$ V to $248.215$ V but for the calculation of power flow equations it has always been $310.2687$ V. Hence, the references are unchanged when a step change in $V_g$ is considered. The responses of the $\tt{dq}$ currents $i_{\tt dq}$ and angle $\varphi$ are shown in Fig. \ref{Vgg-step}. It can be observed that after small transients, both ${\tt dq}$ currents and angle $\varphi$ are quickly restored to their reference values.}
%
\subsection{Experimental results}
\label{sec5}
\begin{figure*}
  \centering
  \includegraphics[scale=0.6]{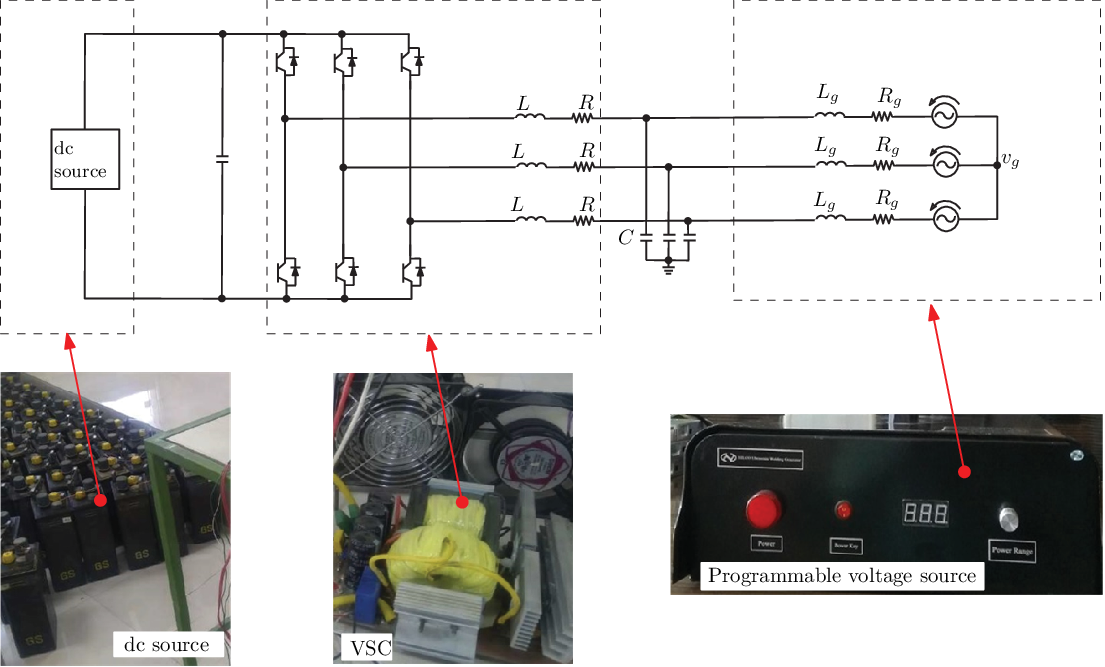}
  \vspace{0cm}
  \caption{Schematic of experimental setup with pictures of hardware components.}\label{setup}
\end{figure*}

In this section experimental results are illustrated to further validate the effectiveness of the proposed solution. A schematic of the experimental setup with pictures of the hardware components is given in Fig. \ref{setup}. The experimental parameters are the same adopted for simulations---see Tabs. \ref{tab1} and \ref{tab2}. Moreover, the setting of experiment scheme and test scenarios are same with them of the simulation.

The real time control of the VSC is implemented via a STM32F407VGT6 microcontroller where, as for simulations, the conversion from the $\tt abc$ to the $\tt dq$ frame is realized via the angle $\vartheta$ determined by the proposed adaptive PLL. The overall architecture employs voltage and current measurements obtained via dedicated sensors. In particular, the scheme described in Fig. \ref{fig0} requires the read of the current of the output filter via the LEM-100P current sensor, its transformation in $\tt{dq}$ coordinates, and the comparison with the corresponding reference value, a procedure that allows for the calculation of the current regulation error to be fed to the microcontroller. On the other hand, the output voltage of the converter is read by resistive division and compared with the variable reference value. By using this approach to measure the voltage, it is possible to isolate and amplify the sampled voltage (in the range of $\pm 250$ mv) through AMC1301, and then bring it to the range of 0 to 3.3 through OPAMP-based signal conditioning stage. An isolated DC-DC converter manufactured by CUI INC named PDQE15-Q24-S15-D with an output of $15$ V and a current of $1$ A is adopted to provide isolated voltage for feeding electronic circuits.  The LM2575 switching regulator is used in this board to provide $5$ V power supply for electronic circuits. Also, for hardware overcurrent protection, the amplified voltage is compared with the final current limit through a comparator TLC372 and, in case of overcurrent, the OCP command cuts off the input of the board. The IGBTs of the employed VSC are characterized by a voltage of $1200$ V and a current of $40$ A. The HCPL3120 gate drives are used to turn on the switches with a voltage of $15$ V and turn them off with a voltage of $-5$ V. The maximum input voltage is considered to $1000$ V and the time is equal to $16$ $\mu$s, and a switching frequency of about $30$ kHz. The battery set is the GS-Yuasa HS-200 series.

\begin{figure}[!ht]
  \centering
  \includegraphics[scale=0.35,trim=130 0 0 0,clip]{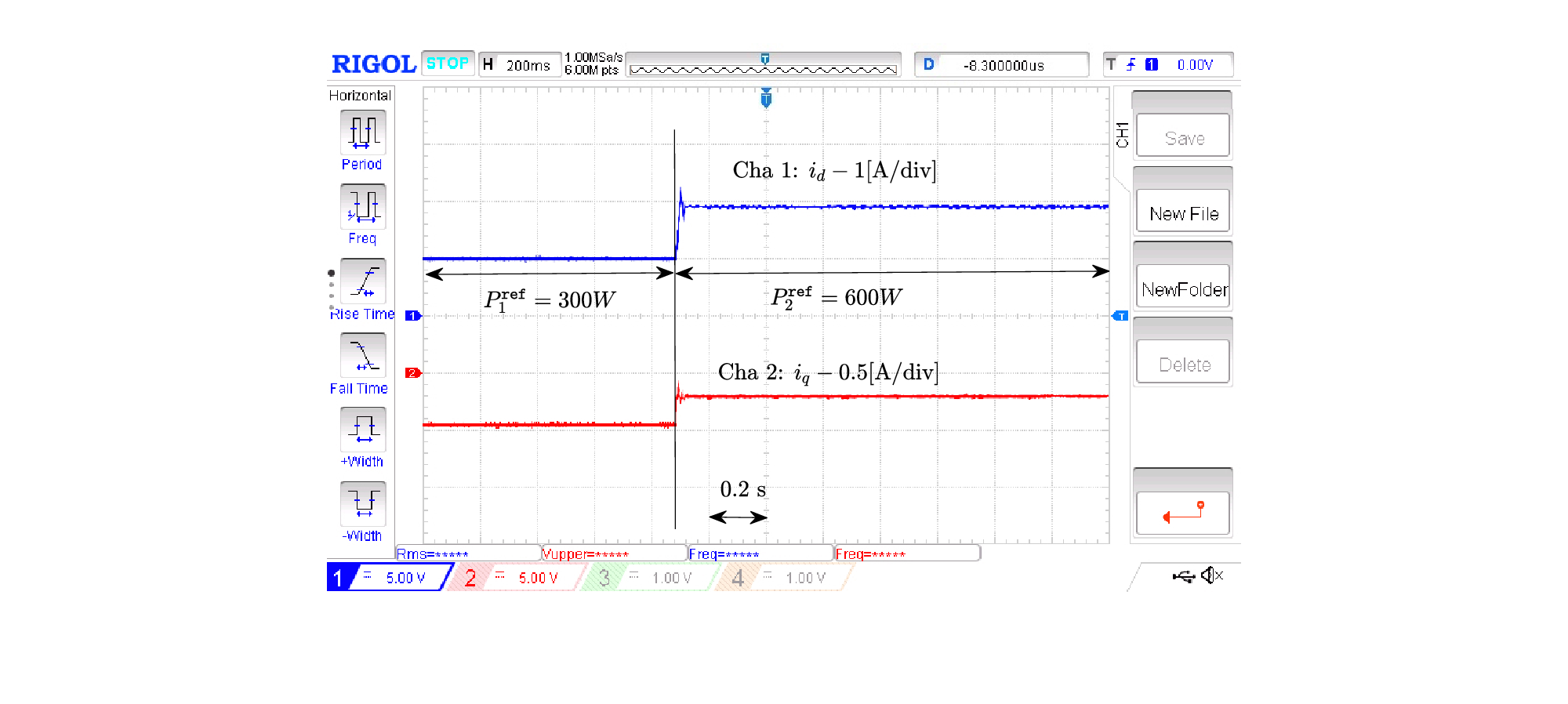}
   \vspace{-1cm}
  \caption{Responses of ${\tt dq}$ currents, facing an active power step variation from $P^\mathrm{ref}_1$ to $P^\mathrm{ref}_2$.}\label{fig1}
\end{figure}

\begin{figure}[!ht]
  \centering
  \includegraphics[scale=0.35,trim=130 0 0 0,clip]{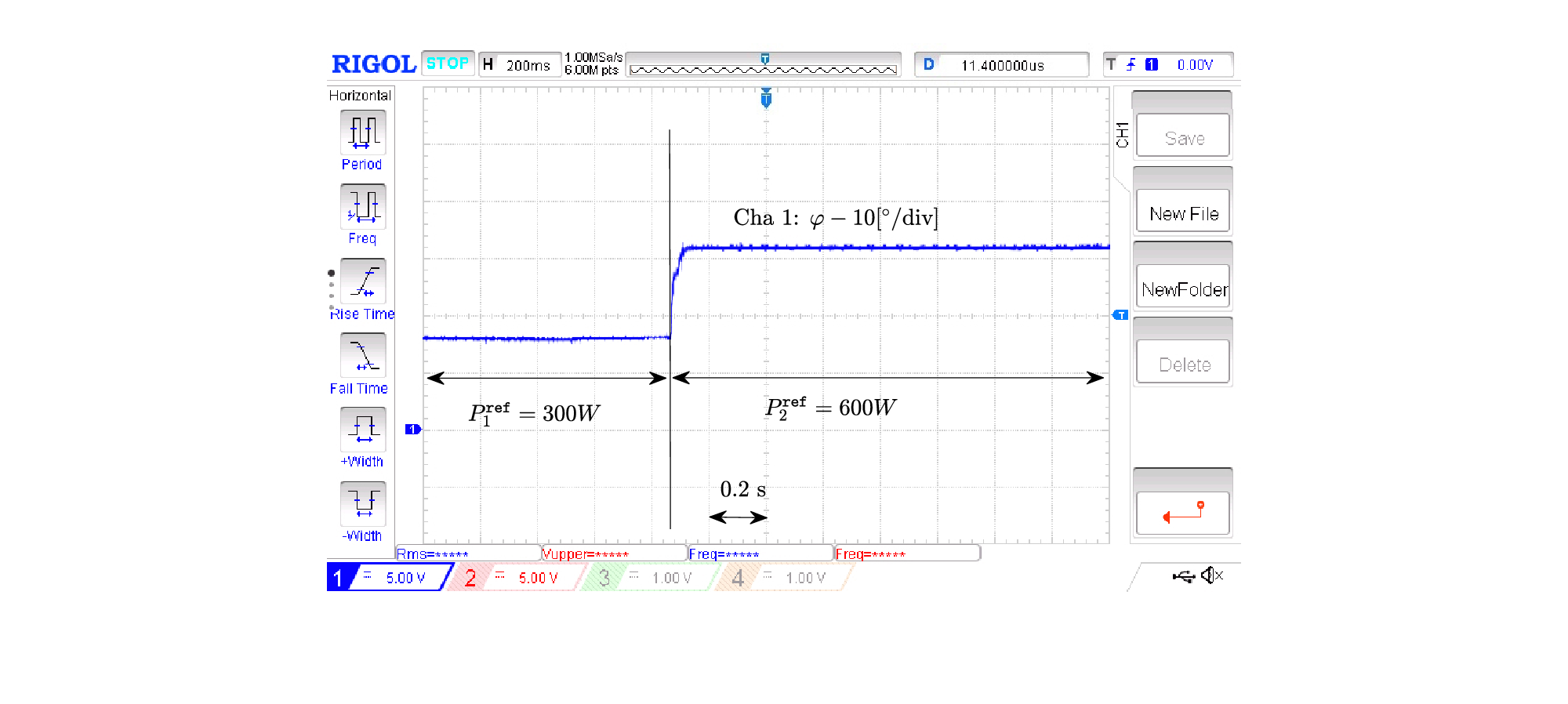}
   \vspace{-1cm}
  \caption{Response of angle $\varphi$, facing an active power step variation from $P^\mathrm{ref}_1$ to $P^\mathrm{ref}_2$.}\label{fig2}
\end{figure}

As for simulations, we consider a step change in the desired power, stepping from $P^\mathrm{ref}_1$ to $P^\mathrm{ref}_2$, assuming that $V^\mathrm{ref}$ is fixed at $380~\text{V}$. The experimental results are shown in Figs. \ref{fig1} and \ref{fig2}. It is observed that the corresponding references are tracked within $0.2~\text{s}$, a result that is consistent to the results observed in simulations.

\begin{figure}[!ht]
  \centering
  \includegraphics[scale=0.35,trim=130 0 0 0,clip]{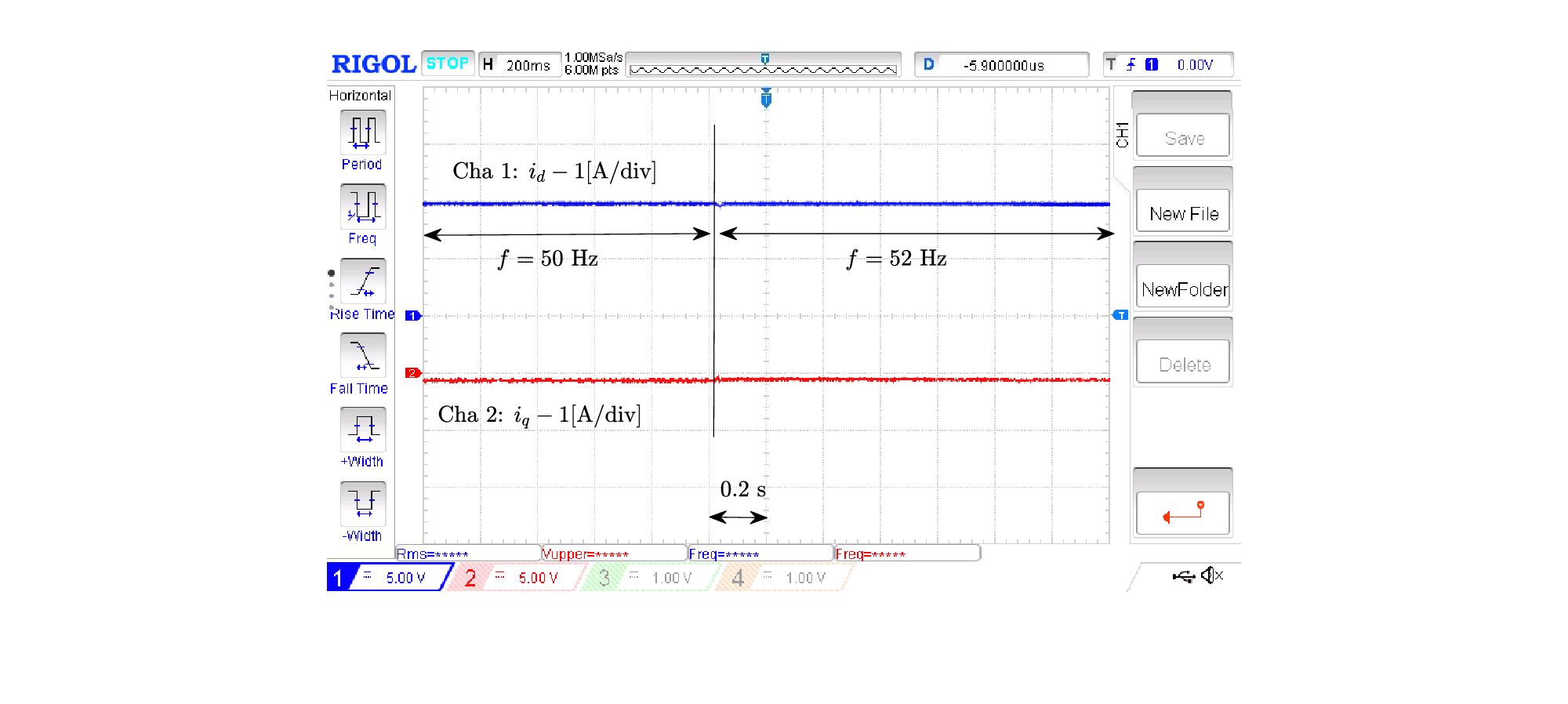}
   \vspace{-1cm}
  \caption{Responses of {\tt dq} currents, facing a frequency step variation of $2$ Hz.}\label{fig3-7}
\end{figure}

\begin{figure}[!ht]
  \centering
  \includegraphics[scale=0.35,trim=130 0 0 0,clip]{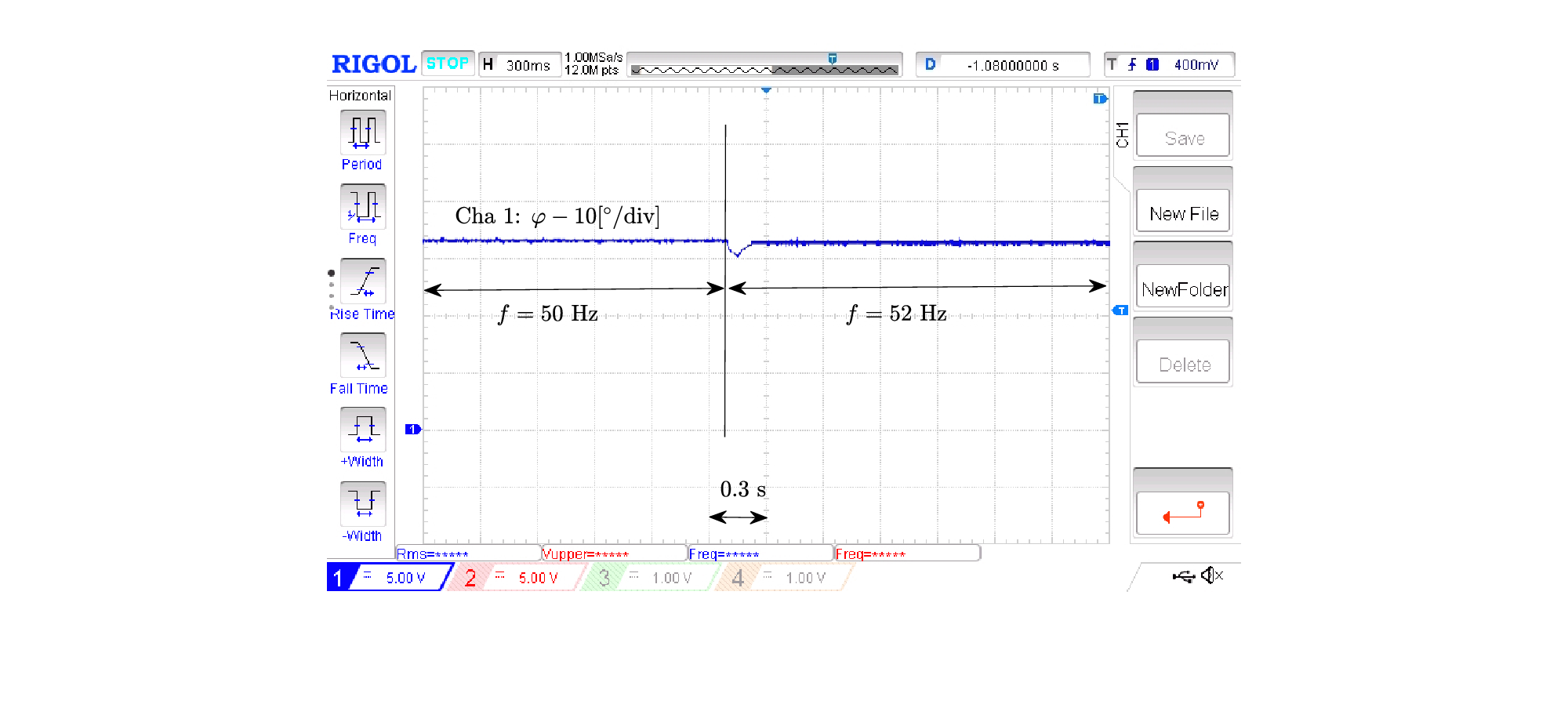}
   \vspace{-1cm}
  \caption{Response of angle $\varphi$, facing a frequency step variation of $2$ Hz.}\label{fig3}
\end{figure}

\begin{figure}[!ht]
  \centering
  \includegraphics[scale=0.35,trim=130 0 0 0,clip]{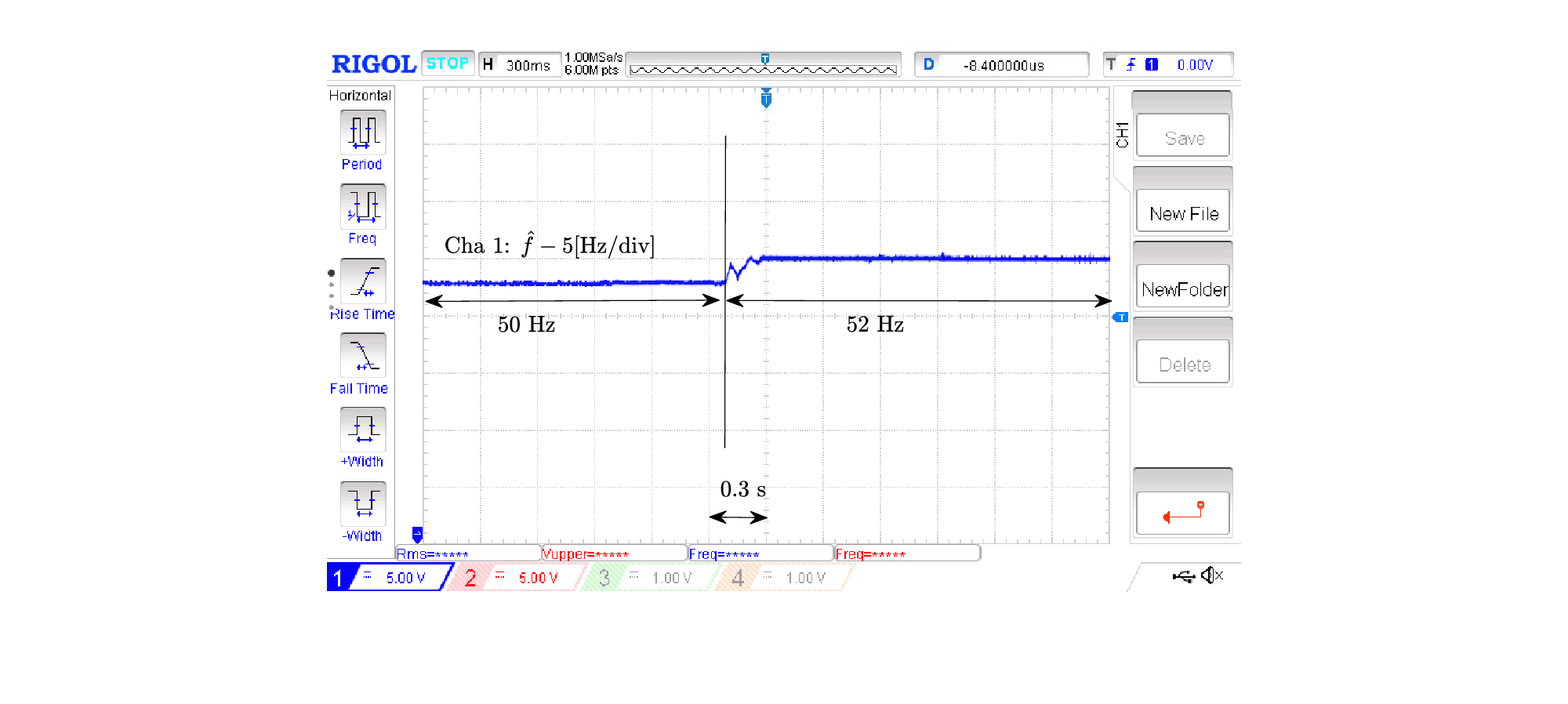}
   \vspace{-1cm}
  \caption{Response of the estimate of $f$, facing a frequency step variation of $2$ Hz.}\label{w-exp}
\end{figure}

Then, the desired power is set to $600~\text{W}$, with voltage  $V^\mathrm{ref}$ maintained at $380~\text{V}$. A change in $f$ is thus imposed, stepping from $50$ Hz to $52$ Hz. This contributes to assess the performance of designed observer under frequency  variation. It is seen from Figs. \ref{fig3-7} and \ref{fig3} that although there is an abrupt variation of the frequency, its perturbed value is rapidly estimated via the observer, ensuring that the $\tt{dq}$  current control objective is still guaranteed after reasonable transients. The estimate of $f$ is illustrated in Fig. \ref{w-exp}.

\begin{figure}[!ht]
  \centering
  \includegraphics[scale=0.35,trim=130 0 0 0,clip]{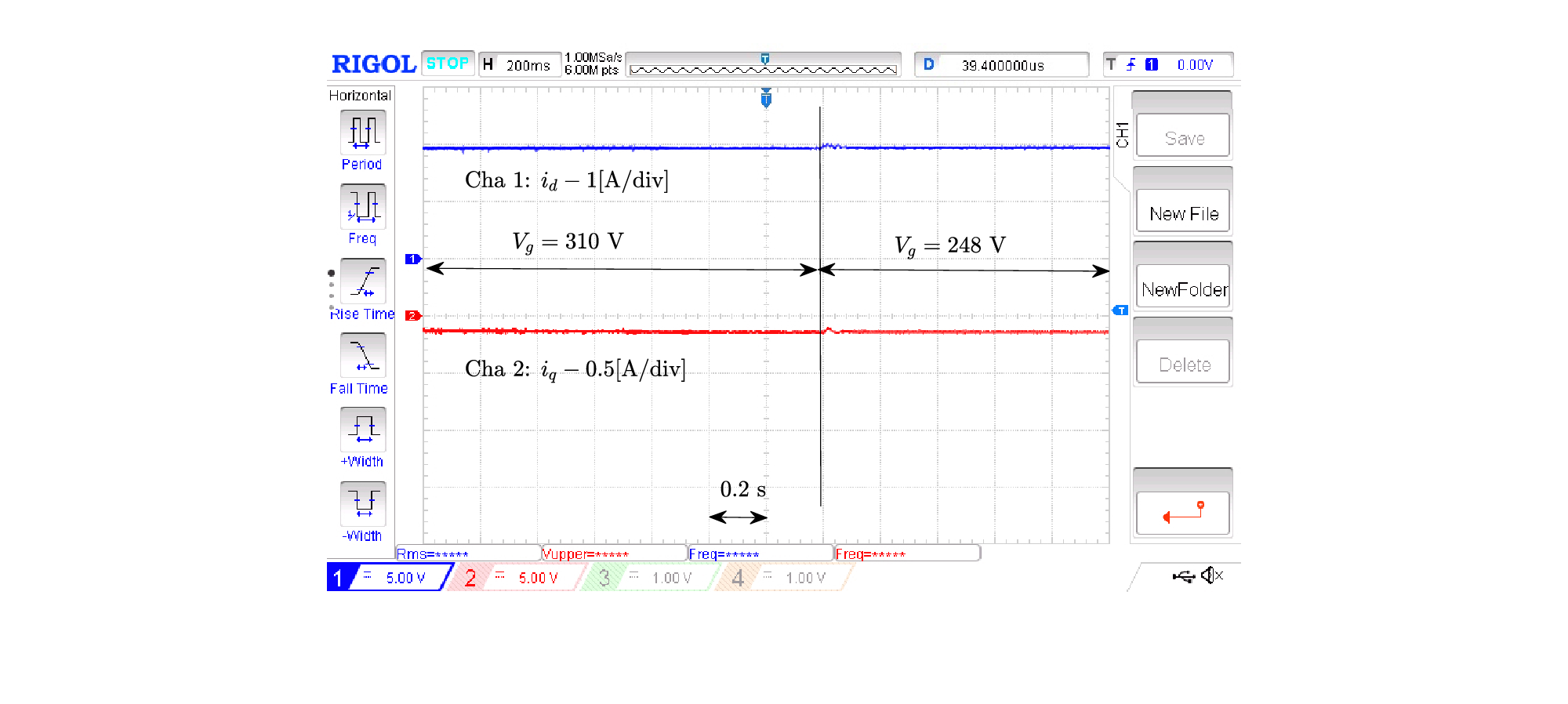}
   \vspace{-1cm}
  \caption{Responses of ${\tt dq}$ currents, facing a voltage amplitude step variation $20 \%$.}\label{fig5}
\end{figure}

\begin{figure}[!ht]
  \centering
  \includegraphics[scale=0.35,trim=130 0 0 0,clip]{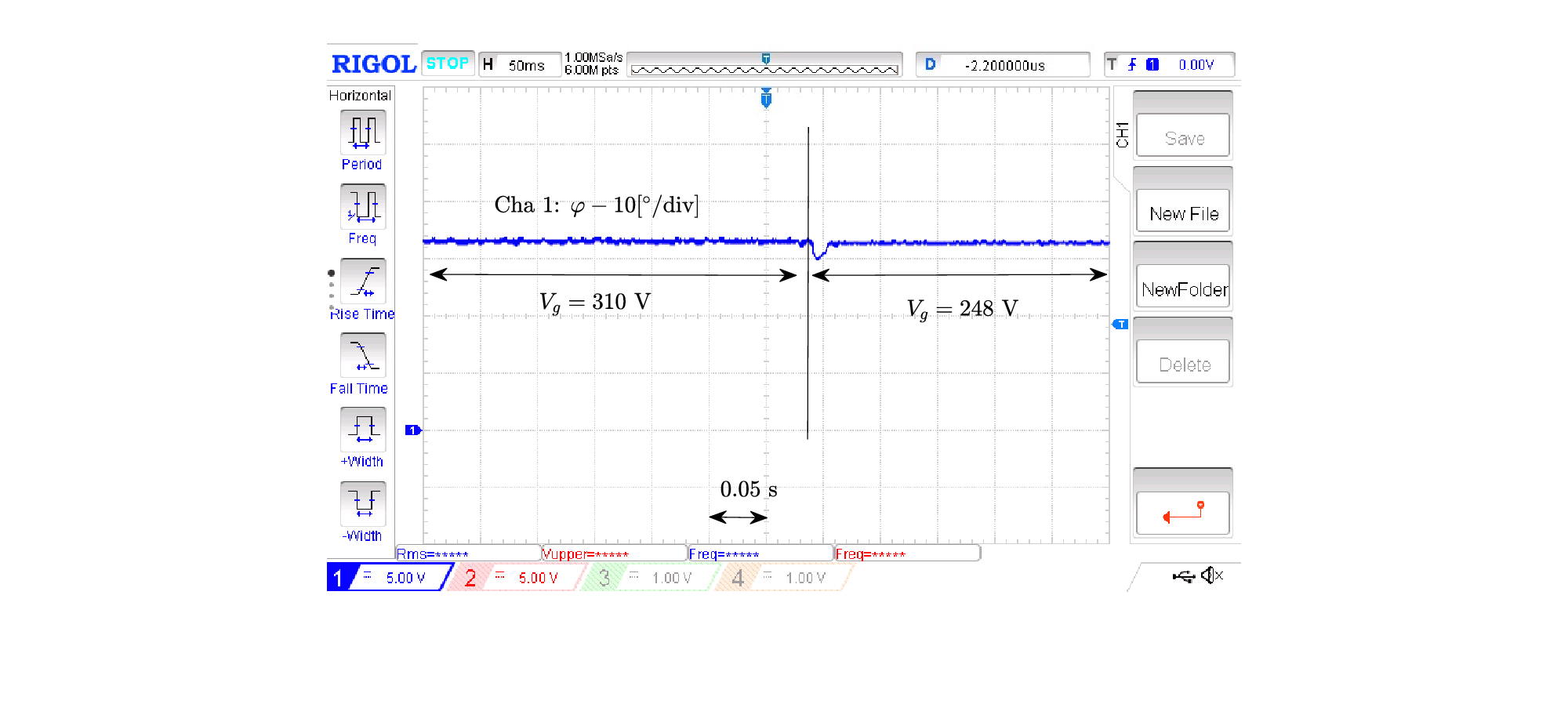}
   \vspace{-1cm}
  \caption{Response of angle $\varphi$, facing a voltage amplitude step variation of $20 \%$.}\label{fig6}
\end{figure}

\begin{figure}[!ht]
  \centering
  \includegraphics[scale=0.35,trim=130 0 0 0,clip]{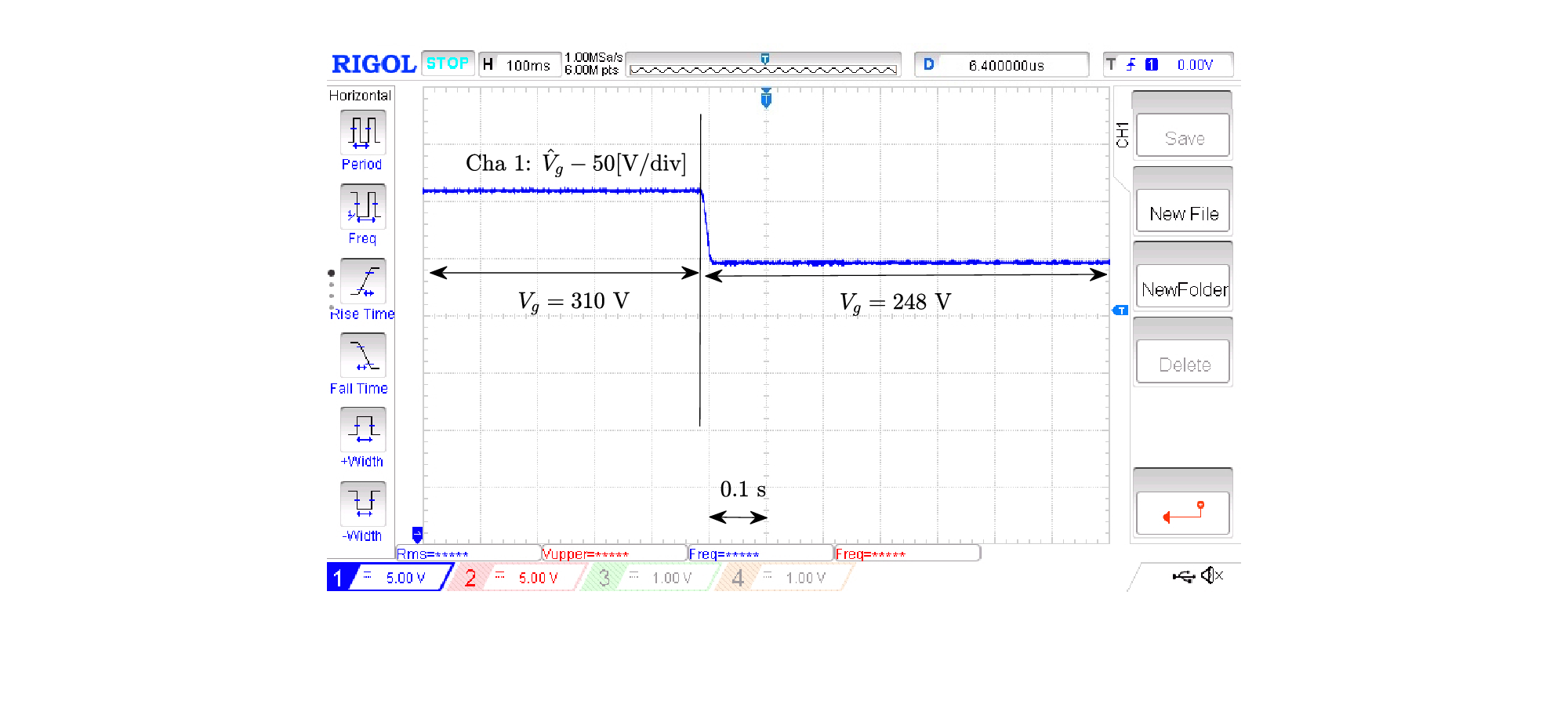}
   \vspace{-1cm}
  \caption{Response of the estimate of $V_g$.}\label{Vg-exp}
\end{figure}

Finally, we validate the ability of the observer to accurately reconstruct the information of the grid voltage amplitude $V_g$ in case of a voltage perturbation. A step change in $V_g$ is therefore considered, in which its $20\%$ variation is faced. This test case is same with the simulation study. The responses of the $\tt{dq}$ currents $i_{\tt dq}$ and angle $\varphi$ are shown in Figs. \ref{fig5} and \ref{fig6}. It is seen that they are quickly restored to their desired values after reasonably small transients, demonstrating robust performance of the solutions. The estimate of $V_g$ is illustrated in Fig. \ref{Vg-exp}, where it is shown that the perturbed value of the voltage is rapidly estimated.

%
\section{Concluding Remarks}\label{sec6}
In this paper we addressed the problem of synchronization of grid-connected VSCs. The work is built upon the observer-based adaptive PLL solution proposed in~\cite{ZONetal_aca} and provided experimental validation of this innovative approach, via the realization of a comprehensive series of experiments.
Through these experiments, we bridge the gap between theoretical expectations and practical outcomes, finally demonstrating the adaptability and robustness of our design. In the future work, we intend to introduce the estimated terms to the calculation of power flow equations. Note that the observer dynamics obviously affects the control performance. In this version, we only consider the nominal conditions.

\bibliographystyle{IEEEtran}
\bibliography{TPE23}

\end{document}